\documentclass[prl,aps,amsfonts,twoside,twocolumn,amssymb,superscriptaddress]{revtex4-1}

\usepackage[latin1]{inputenc}
\usepackage[english]{babel}
\usepackage{graphicx}
\usepackage{amsmath}
\usepackage{units}
\usepackage{color}
\usepackage{ulem}
\usepackage{soul}
\usepackage{float}
\usepackage[caption=false]{subfig} 
\usepackage{blindtext}

\setcitestyle{super}

\newcommand{\abs}[1]{\lvert#1\rvert}

\newcommand{\estiset}{\mathcal{E}}
\newcommand{\least}{\check{X}}
\newcommand{\most}{\hat{X}}
\newcommand{\hi}[1]{\hat{#1}}
\newcommand{\lo}[1]{\check{#1}}
\newcommand{\alphakg}{\chi_\text{kg}}

\begin{document}

\title{Implementation of Continuous-Variable Quantum Key Distribution with Composable and One-Sided-Device-Independent Security Against Coherent Attacks}

\author{Tobias Gehring}
\affiliation{Max-Planck-Institut f\"ur Gravitationsphysik
(Albert-Einstein-Institut) and\\ Institut f\"ur Gravitationsphysik, Leibniz Universit\"at Hannover, Callinstraße 38, 30167 Hannover,
Germany}
\affiliation{Department of Physics, Technical University of Denmark, Fysikvej, 2800 Kgs.\,Lyngby, Denmark}
\author{Vitus H\"andchen}
\affiliation{Max-Planck-Institut f\"ur Gravitationsphysik
(Albert-Einstein-Institut) and\\ Institut f\"ur Gravitationsphysik, Leibniz Universit\"at Hannover, Callinstraße 38, 30167 Hannover,
Germany}
\affiliation{Institut f\"ur Laserphysik und Zentrum f\"ur Optische Quantentechnologien, Universit\"at Hamburg, Luruper Chaussee 149, 22761 Hamburg, Germany}
\author{Jörg Duhme}
\affiliation{Institut f\"ur Theoretische Physik, Leibniz Universit\"at Hannover, Appelstraße 2, 30167 Hannnover, Germany}
\author{Fabian Furrer}
\affiliation{Department of Physics, Graduate School of Science, University of Tokyo, 7-3-1 Hongo, Bunkyo-ku, Tokyo, Japan, 113-0033}
\author{Torsten Franz}
\affiliation{Institut f\"ur Theoretische Physik, Leibniz Universit\"at Hannover, Appelstraße 2, 30167 Hannnover, Germany}
\affiliation{Institut f\"ur Fachdidaktik der Naturwissenschaften, Technische Universit\"at Braunschweig, Bienroder Weg 82, 38106 Braunschweig, Germany}
\author{Christoph Pacher}
\affiliation{Digital Safety \& Security Department, AIT Austrian Institute of Technology GmbH, 1220 Vienna, Austria}
\author{Reinhard F. Werner}
\affiliation{Institut f\"ur Theoretische Physik, Leibniz Universit\"at Hannover, Appelstraße 2, 30167 Hannnover, Germany}
\author{Roman Schnabel}
\email[corresponding author: ]{roman.schnabel@physnet.uni-hamburg.de}
\affiliation{Max-Planck-Institut f\"ur Gravitationsphysik
(Albert-Einstein-Institut) and\\ Institut f\"ur Gravitationsphysik, Leibniz Universit\"at Hannover, Callinstraße 38, 30167 Hannover,
Germany}
\affiliation{Institut f\"ur Laserphysik und Zentrum f\"ur Optische Quantentechnologien, Universit\"at Hamburg, Luruper Chaussee 149, 22761 Hamburg, Germany}

\begin{abstract}
Secret communication over public channels is one of the central pillars of a modern information society.
Using quantum key distribution this is achieved without relying on the hardness of mathematical problems which might be compromised by improved algorithms or by future quantum computers.
State-of-the-art quantum key distribution requires composable security against coherent attacks for a finite number of distributed quantum states as well as robustness against implementation side-channels. 
Here, we present an implementation of continuous-variable quantum key distribution satisfying these requirements.
Our implementation is based on the distribution of continuous-variable Einstein-Podolsky-Rosen entangled light.
It is one-sided device independent, which means the security of the generated key is independent of any memoryfree attacks on the remote detector.
Since continuous-variable encoding is compatible with conventional optical communication technology, our work is a crucial step towards practical implementations of quantum key distribution with state-of-the-art security based solely on telecom components.
\end{abstract}

\maketitle

Using a quantum key distribution (QKD) system the communicating parties employ a cryptographic protocol that cannot be broken, neither by todays nor by future technology~\cite{Gisin2002,Scarani2009}. The security of the key distributed by such a system is guaranteed on the basis of quantum theory by a mathematical proof, which has to consider the most sophisticated (quantum) attacks on the quantum channel, so-called `coherent attacks'. 
Furthermore, security has to be established in a `composable' fashion, which means that if the distributed key is used in another secure protocol (like one-time-pad encryption), it remains secure in the composition of the two protocols~\cite{BenOr2005,Renner2005}. To make a security proof applicable to actual implementations, it is important to include all effects due to the finite number of distributed quantum states. Additionally, the security proof has to model the source and the detectors correctly to prevent possible `side-channels', including those which may only be discovered in the future.
\begin{figure}[ht!!!!!!!!!!!!!!!!!!!!]
    \includegraphics[width=8.5cm]{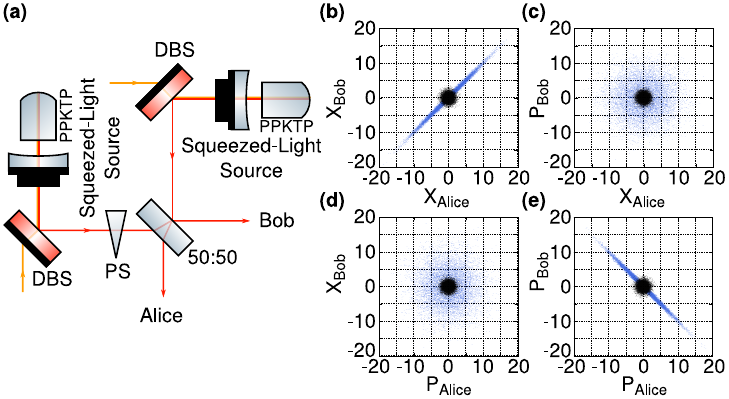}
    \caption{\textbf{Einstein-Podolsky-Rosen entanglement source for CV QKD.} (a) The source consists of two continuous-wave squeezed vacuum beams, generated by type I parametric down-conversion at $1550$\,nm (red), which are superimposed at a balanced beam splitter with a relative phase of $\tfrac{\pi}{2}$. Yellow beam: $775$\,nm pump field, DBS: Dichroic beam splitter, PS: Phase shifter. (b)-(e) Correlations between Alice's and Bob's data, measured by balanced homodyne detection in either the amplitude ($X$) or phase ($P$) quadrature. The data is normalized to the noise standard deviation of a vacuum state. Blue: Einstein-Podolsky-Rosen entangled state used for QKD. Black: Reference measurement of zero-point fluctuations of the ground state (vacuum).}
    \label{fig:quadrature entanglement}
\end{figure}

Theoretically, an elegant way to deal with imperfect sources and detectors and therefore with side-channels of the implementation, is to make a proof completely device independent~\cite{Acin2007}. 
The found secret key rates are, however, very low so far and an implementation requires at least a detection-loophole-free Bell test, which has not been achieved in a QKD implementation so far due to inefficient detectors and photon loss in the quantum channel~\cite{Acin2007}. 
The idea of removing assumptions on devices can nevertheless be realized partially. 
For instance, measurement-device-independent (MDI) QKD relies only on assumptions about the sources, located at the honest communicating parties, Alice and Bob, but not about the detectors which can be in control of the eavesdropper~\cite{Braunstein2012,Lo2012,Pirandola2015}. 
While in MDI QKD the devices of Alice and Bob have to be trusted to fulfill the assumptions, it has recently been shown that QKD is even possible when the device of one of the honest parties is untrusted~\cite{Tomamichel2011,Branciard2012,Tomamichel2013}.
For discrete variables the security of this one-sided device independent (1sDI) scheme has been analyzed under the assumption on the untrusted device to be memoryless, and similar secret key rates have been obtained as in QKD implementations with trusted devices only~\cite{Tomamichel2011,Branciard2012,Wang2013}.
Using continuous variables (CV) 1sDI QKD has been recently proven secure for collective attacks and infinitely many quantum state distributions~\cite{Walk2014} as well as with finite-size, composable security against coherent attacks under the same assumption of a memoryless untrusted device~\cite{Furrer2012}.

So far experimental continuous-variable implementations were only guaranteed to be secure against so-called `collective attacks'~\cite{Cerf2001,Weedbrook2012,Lodewyck2007,Jouguet2013}.
While this class of attacks already allows an eavesdropper to possess a quantum memory, all quantum states are attacked identically using a collective Gaussian operation. 
Although Gaussian collective attacks are in the limit of an infinite number of distributed quantum states as strong as coherent attacks, it is currently not known whether this holds for a realistic finite key length protocol.
For collective attacks a transmission distance of $80$\,km was achieved with a finite number of distributed quantum states using Gaussian modulated coherent states\cite{Leverrier2010,Jouguet2013}.
Previous proofs did also find composable security against coherent attacks for continuous variables~\cite{Renner2009,Leverrier2013} but only for an unrealistically large number of distributed quantum states. 

Here, we report a continuous-variable QKD implementation that generates a finite and composable key which is secure against coherent attacks and whose security is furthermore 1sDI under memoryless assumption.
Theoretically, the security of our implemented protocol is based on an extension of the security proof in Ref.~\citenum{Furrer2012} including measurement flaws in the trusted detector. 
Our implementation is based on Gaussian Einstein-Podolsky-Rosen (EPR) entangled light and homodyne detection as considered in the security proof. 
An optimized, highly efficient error reconciliation algorithm was developed to enable the generation of the secret key.


\section{Results}

\subsection{Robustness against Implementation Side-Channels}

The 1sDI QKD implementation presented here is very robust against implementation side-channel attacks. It is secure against memoryfree attacks performed on Bob's untrusted detector, i.e.\ attacks that are independent on Bob's previous measurement outcomes. This includes recently proposed attacks on the intensity of the local oscillator~\cite{Ma2013,Ma2014}, calibration attacks of the shot-noise reference~\cite{Jouguet2013a,Kunz-Jacques2015}, wavelength attacks on the homodyne beam splitter~\cite{Ma2013a,Huang2013} and saturation attacks on the homodyne detector's electronic circuit~\cite{Qin2013}. 
Furthermore it is secure against trojan-horse attacks on the source which usually threaten electro-optical modulators commonly used in Gaussian-modulation QKD protocols~\cite{Jain2014,Lo1999}.
Placing the EPR source at Alice's station and assuming that her station is private and inaccessible to the eavesdropper by other means than the quantum channel~\cite{Braunstein2012}, prevents exploiting side-channels related to the local oscillator used by Alice's trusted detector as the eavesdropper simply has no way of accessing it.
Saturation attacks on Alice's homodyne detector are directly prevented by the security proof which includes an upper and lower bound for measurement outcomes~\cite{Furrer2012,Qin2013}.

\subsection{Einstein-Podolsky-Rosen Source}

Our implemented protocol uses two continuous-wave optical light fields whose amplitude and phase quadrature amplitude modulations were mutually entangled~\cite{Furusawa1998}, produced by a source which is the only component in the setup which is not compatible with existing telecommunication components. Using Einstein-Podolsky-Rosen entanglement as a resource makes our protocol a CV equivalent of the BBM92 protocol for discrete variables~\cite{Bennett1992}.
The schematic of the experimental setup is illustrated in Fig.~\ref{fig:quadrature entanglement}(a). Two squeezed-light sources~\cite{Eberle2010,Mehmet2011}, each composed of a nonlinear PPKTP crystal and a coupling mirror, were pumped with a bright pump field at $775$\,nm (yellow) to produce two squeezed vacuum states at the telecommunication wavelength of $1550$\,nm (red). The two squeezed vacua, both exhibiting a high squeezing of more than $10$\,dB, were superimposed at a balanced beam splitter with a relative phase of $\pi/2$, thus generating Einstein-Podolsky-Rosen entanglement~\cite{Furusawa1998}. One of the outputs of the beam splitter was kept by Alice, while the other was sent to Bob. The technical details of the source, including the locking scheme, were characterized in Ref.~\citenum{Eberle2013}.

\begin{figure}[h]
    \includegraphics[width=8.5cm]{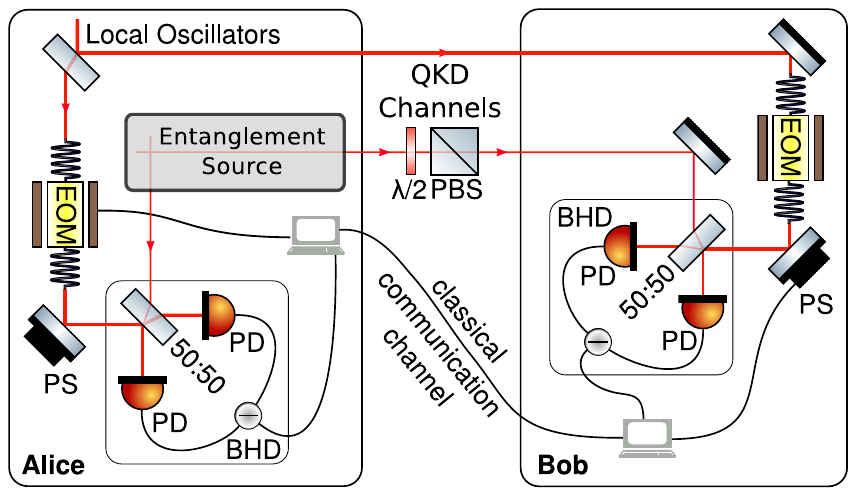}
    \caption{\textbf{Implementation of Alice's and Bob's QKD receivers}. Both parties used balanced homodyne detection (BHD) to measure their part of the quadrature entangled state. The measured quadrature angle was controlled by a computer via a fast fiber-coupled electro-optical modulator (EOM). To make sure that Alice and Bob switched between the same orthogonal quadratures, a phase shifter (PS) was employed to compensate slow phase drifts (see Methods). Optical losses of the transmission channel to Bob were modelled by a variable attenuator consisting of a half-wave plate ($\lambda/2$) and a polarizing beam splitter (PBS). The measurement rate was $100$\,kHz. PD: Photo Diode.}
    \label{fig:detectors}
\end{figure}

Figures~\ref{fig:quadrature entanglement}(b)-(e) show the distribution of measurement outcomes obtained by the two parties measuring either the amplitude ($X$) or phase ($P$) quadrature of their respective light field with balanced homodyne detection. Each measurement outcome is truly random since it stems from parametrically amplified zero-point fluctuations. When both parties simultaneously measure either $X$ or $P$ the strong correlations between their outcomes are clearly visible (Fig.~\ref{fig:quadrature entanglement} (b) and (e)). If the two parties measure different quadratures instead, the measurement outcomes are uncorrelated (Fig.~\ref{fig:quadrature entanglement}(c) and (d)). The strength of the correlations of Alice's and Bob's measurement for the same quadratures, which is related to the initial squeezing strength, is a central parameter in our QKD protocol and enters the key length computation directly in the form of an average distance $d_\text{pe}$, introduced below. 

A schematic of the experimental QKD setup is shown in Fig.~\ref{fig:detectors}. The entanglement source was located at Alice's station and the local oscillators used for homodyne detection of the two entangled modes were generated locally at her station as well. While this assured that Alice's local oscillator was inaccessible to an eavesdropper, Bob's local oscillator was sent from Alice to Bob via a free-space channel. Both local oscillators had a power of $10$\,mW each. Implementation details can be found in the Methods.

\subsection{Precise Steps of the QKD Protocol}

Preliminaries; Alice and Bob use a pre-shared key to authenticate the classical communication channel for post processing~\cite{Stinson1994,Gemmell1994}. Furthermore, Alice and Bob negotiate all parameters needed during the protocol run and Alice performs a shot-noise calibration measurement by blocking the signal beam input of her homodyne detector.

Measurement Phase; Alice prepares an entangled state using her Einstein-Podolsky-Rosen source and sends one of the outputs to Bob along with a local oscillator beam. Both Alice and Bob choose, randomly and independently from each other, a quadrature $X$ or $P$, which they simultaneously measure by homodyne detection of their light fields. The outcome of this measurement is called a sample. This step is repeated until $2N$ samples have been obtained. 

Sifting; Alice and Bob announce their measurement bases and discard all samples measured in different quadratures.

Discretization; The continuous spectrum of the measurement outcomes is discretized by the analog-to-digital converter (ADC) used to record the measurement. During the discretization step Alice and Bob map the fine grained discretization of their remaining samples caused by the ADC to a coarser one consisting of consecutive $2^d$ bins. In the interval $[-\alpha,\alpha ]$ a binning with equal length is used, which is complemented by two bins $(-\infty,-\alpha)$ and $(\alpha,\infty)$. The parameter $\alpha$ is used to include the finite range of the homodyne detectors into the security proof. 

Channel Parameter Estimation; The secret key length is calculated using the average distance between Alice's and Bob's samples. To estimate it, the two parties randomly choose a common subset of length $k$ from the sifted and discretized data, $X_A^\text{pe}$ and $X_B^\text{pe}$, respectively, which they communicate over the public classical channel. Using these, they calculate
\begin{equation}
    \label{eqn:pe test}
    d_\text{pe}(X_A^\text{pe}, X_B^\text{pe}) = \frac{1}{k}\sum_{\mu=1}^k \abs{(X_A^\text{pe})_\mu - (X_B^\text{pe})_\mu}\ ,
\end{equation}
and abort if it exceeds a threshold agreed on in the preliminaries step.

Error Reconciliation; Bob corrects the errors in his data to match Alice's using the hybrid error reconciliation algorithm described below. Afterwards, Alice and Bob confirm that the reconciliation was successful. 

Calculation of Secret Key Length; Using the results from the channel parameter estimation and considering the number of published bits during error reconciliation, Alice and Bob calculate the secret key length $\ell$ according to the presented secret key length formula in the Methods. If the secret key length is negative, they abort the protocol.

Privacy Amplification; Alice and Bob apply a hash function which is randomly chosen from a two-universal family~\cite{Carter1979}, to their corrected strings to produce the secret key of length $\ell$.

\subsection{Assumptions of the Security Proof}

The assumptions of the security proof on our implementation are the following: 1) Alice's station is a private space~\cite{Braunstein2012} and Bob's station is isolated, i.e.\ neither Bob's measurement choice nor his measurement results are leaking his station. 2) The energy of Alice's mode of the EPR state is bounded which allows Alice to determine the probability for measuring a quadrature amplitude value exceeding the parameter $\alpha$.  3) Alice switches her homodyne detector randomly between two orthogonal quadratures ($X$ and $P$) with $50$\,$\%$ probability. 4) Bob is choosing randomly between two measurements that are assumed to be memoryless. 5) The phase noise present in Alice's measurement is Gaussian distributed with variances $V_X$ and $V_P$ for the amplitude and phase quadrature, respectively.

The first assumption is natural to (almost) all QKD implementations. 
The second one is assured in our implementation by placing the EPR source into Alice's station.
For the third and fourth assumptions two independent quantum random number generators located at Alice's and Bob's stations were employed.
For implementation details we refer to the Methods. While Bob is choosing randomly between two measurements, it is not required that they are orthogonal quadrature measurements. Since the security of the key is independent of the actual measurements, an eavesdropper may temper with the local oscillator sent to Bob.
In an experimental implementation phase noise is unavoidable, hence the security proof of Ref.~\citenum{Furrer2012} has been extended, see Methods for details.
We characterized the phase noise in our implementation before the run of the protocol, showed that the quadratures are indeed Gaussian distributed and determined the variances to $V_X = V_P \approx (0.46^\circ \pm 0.01^\circ)^2$.
Details are given in the Methods. 
Thus, our implementation fulfills all requirements of the security proof and the generated key by the above protocol is $\epsilon$-secure against coherent attacks, where $\epsilon$ is the so-called composable security parameter.

\subsection{Error Reconciliation Protocol}

Important for a high key rate is an error reconciliation protocol which has an efficiency close to the Shannon limit. Since in our CV QKD protocol the discretized sample values are non-binary and follow a Gaussian distribution, error reconciliation codes with high efficiency and low error rate are more difficult to achieve than for discrete-variable protocols with uniformly distributed binary outcomes~\cite{Lodewyck2007}. To solve the problem, we designed a two-phase error reconciliation protocol which can exploit the non-uniform distribution efficiently. First the $d_1$ least significant bits of each sample are sent to Bob. Since these bits are only very weakly correlated this step works with an efficiency very close to the Shannon limit.
In a second step Alice and Bob use a non-binary low density parity check (LDPC) code over the Galois field $\text{GF}(2^{d_2})$ to correct the $d_2=d-d_1$ most significant bits. $d_1$, $d_2$, as well as the LDPC code were optimized for the different channel conditions and the actually employed code was determined using the $k$ revealed samples from the channel parameter estimation. More details are given in the Methods.

\begin{figure}[ht]
    \includegraphics[width=8.2cm]{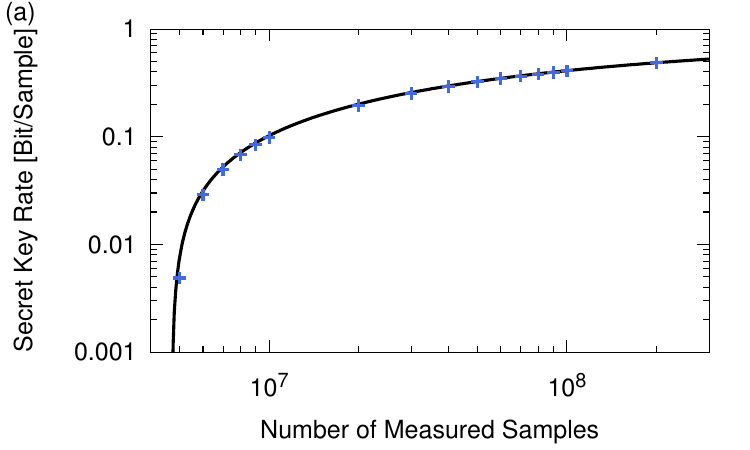}
    \includegraphics[width=8.2cm]{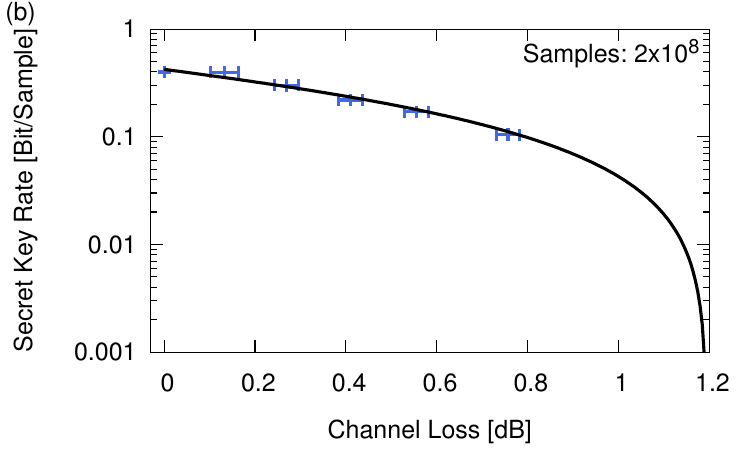}
    \caption{\textbf{Secure key rates achieved by our CV QKD system.} Common parameters: $\alpha=61.6$, $d = 12$, $\epsilon = 2\times 10^{-10}$. (a) Effect of the finite number of distributed quantum states on the secret key rate. The graph shows experimental results (blue points) obtained without the variable attenuator in Bob's arm. The theoretical model (solid line) is included for comparison and was calculated by reconstructing the covariance matrix for $10^8$ samples. (b) Experimentally obtained secure key rate versus optical attenuation in the transmission line to Bob's detector for $2 \times 10^8$ measured samples (blue points). The error bars (standard deviation) are due to the accuracy of the measurement of the optical attenuation. The theoretical model (solid line) was calculated by reconstructing the covariance matrix of the state corresponding to no attenuation ($0$\,dB) and using a reconciliation efficiency of $\beta = 94.3\,\%$.}
    \label{fig:results}
\end{figure}

\subsection{Secret Key Generation}

Figure~\ref{fig:results} shows the experimental results. First we removed the variable attenuator in the transmission line to Bob and executed the protocol for different sample sizes to show the effect of the finite sample size on the secure key rate (Fig.~\ref{fig:results}\,(a), blue points). For each sample size the number of samples $k$ used for channel parameter estimation was optimized before each run of the QKD protocol to yield maximum key length. 
The hybrid error reconciliation had a total efficiency of $\beta = 94.6\,\%$ without a single frame error. While we achieved a positive secret key rate with already $5 \times 10^6$ samples, the secret key rate of $0.485$\,bit/sample achieved for $2 \times 10^8$ samples is close to saturation. The theoretical model, which is the solid line in the figure, is shown for comparison. 

With the variable attenuator in place, we varied the optical loss of the channel to Bob between $0\,\%$ and $16\,\%$ (see Figure~\ref{fig:results}\,(b)), which is equivalent to a fiber length of up to $2.7$\,km when standard telecommunication fibers with an attenuation of $0.2$\,dB/km are used and a coupling efficiency of $95$\,$\%$ is taken into account. By measuring a total of $2\times 10^8$ samples we were still able to achieve a secret key rate of about $0.1$\,bit/sample at an equivalent fiber length of $2.7$\,km ($\approx 0.76$\,dB channel loss). This value, as well as the secret key sizes at the other attenuation values, were achieved by having a very high overall error reconciliation efficiency between $\beta = 94.3\,\%$ and $95.5\,\%$, again without a single frame error.
The theoretical model shown in the figure reveals that even an optical transmission loss of almost $1.2$\,dB between Alice and Bob should be possible. This corresponds to an equivalent distance of about $4.8$\,km, which is already enough to implement CV QKD links with composable 1sDI security against coherent attacks between parties in, for instance, a city's central business district.


\section{Discussion}
In conclusion, we have successfully implemented continuous-variable QKD with composable and 1sDI security against coherent attacks. Along with the exploitation of strong Einstein-Podolsky-Rosen entanglement and a new highly efficient error reconciliation algorithm, the innovation of fast controlled random switching between the two measured quadrature angles with low phase noise made the implementation possible. 
While in our setup Alice and Bob were located on the same optical table, they could in principle be separated and connected by a standard telecommunication fiber (see Methods).

Estimations show that our implementation is limited to about $4.8$\,km. 
Longer distances will be possible by using optical fibres with less loss, or by using reverse reconciliation where about $16$\,km are possible with a similar setup~\cite{Furrer2014}.
Even larger distances, but still remaining secure against coherent attacks in the finite-size regime, require new security proofs since the uncertainty principle employed here yields a secret key rate which does not converge with number of distributed quantum states to the rate achieved for collective attacks and other currently available proofs require an unfeasibly large number of distributed quantum states.
Even more impact will have a further developed proof that keeps all features demonstrated here, but avoids the requirement for an EPR source.
It might be based on Gaussian modulation of coherent states~\cite{Diamanti2015} instead, thus, making 1sDI QKD implemenations possible with composable security against the most general attacks that are solely based on telecommunication components.


\section{Methods}

\subsection{Details of the Experimental Setup}

The measurement rate of our implementation was $100$\,kHz. For each measurement, both Alice and Bob had to choose randomly between the $X$ and $P$ quadrature. The necessary relative phase shifts of $\pi/2$ of the local oscillator with respect to the signal beam were applied to the local oscillator beam by a high-bandwidth fiber-coupled electro-optical phase modulator driven by a digital pattern generator PCI-Express card. Since not only the orthogonality of the measurements is important but also that Alice and Bob measure the same set of quadratures, we compensated slow phase drifts by a phase shifter made of a piezo attached mirror. The error signal for this locking loop was derived by employing an $82$\,MHz single sideband from the entanglement generation~\cite{Eberle2013} which was detected by the homodyne detector. By lowpass filtering the demodulated homodyne signal at $10$\,kHz with a sufficiently high order, the high frequency phase changes from the fiber-coupled phase modulator were averaged over. To make the average independent of the chosen sequence of quadratures we used the following scheme. For a choice of the $X$ quadrature, the phase modulator was first set to a phase of $\pi/2$ during the first half of the $10$\,$\mu$s interval, and then to $0$. For the $P$ quadrature, the phase was first set to $0$ and then to $\pi/2$. Thus, this scheme made sure that the phase did not stay in one quadrature for longer than $10$\,$\mu$s even in the case where one party chose by chance to measure only one quadrature for a while. The measurement was performed synchronously by Alice and Bob in the second half of the interval after $3$\,$\mu$s settling time.

The data acquisition was triggered by the pattern generator and performed by a two channel PCI-Express card at a rate of $256$\,MHz. The $200$ acquired samples per channel were digitally mixed down at $8$\,MHz, lowpass filtered by a $200$-tap finite impulse response filter with a cut-off frequency of $200$\,kHz and down-sampled to one sample. After the total number of samples were recorded the classical post processing of the QKD protocol was performed.

Alice and Bob both employed a local oscillator with a power of $10$\,mW, yielding a dark noise clearance of about $18$\,dB. The efficiency of both homodyne detectors was $98$\,$\%$ (quantum efficiency of the photo diodes $99$\,$\%$, homodyne visibility $99.5$\,$\%$). The pump powers for the two squeezed-light sources were $140$\,mW and $170$\,mW, respectively.

The optical attenuation of the variable attenuator used in Fig.~\ref{fig:results}(b) was measured by determining the strength of the $35.5$\,MHz phase modulation used to lock one of the squeezed-light sources~\cite{Eberle2013} with Bob's homodyne detector. The error bars in the figure are due to the accuracy of this measurement.

While in our implementation both parties were located on the same optical table and the quantum states including the local oscillator for Bob's homodyne detection were transmitted through free space, a separation is in principle possible by using standard telecommunication fibers. To send both the entangled state and the local oscillator to Bob, they could be, for instance, time multiplexed. Using a dedicated fiber for both beams would also be possible. To achieve synchronization between the two parties, a modulated $1310$\,nm beam could be employed which could be send along with the local oscillator by wavelength division multiplexing.

\subsection{Determination Alice's Homodyne Measurement Phase Noise}

\begin{figure}[h]
    \includegraphics[width=8.5cm]{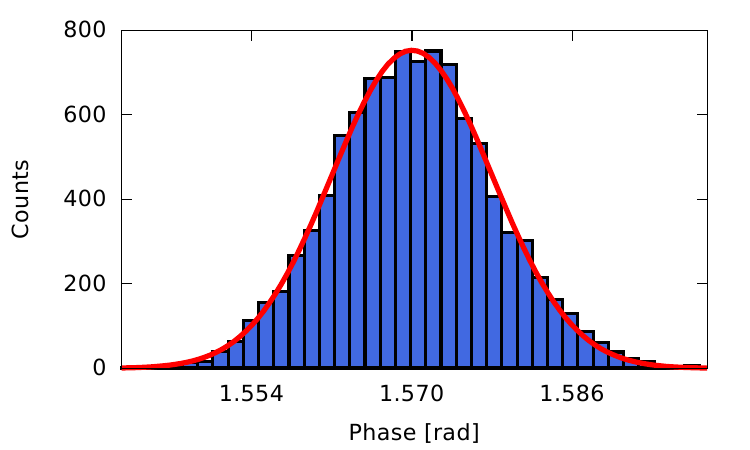}
    \caption{Phase noise measurement result. The standard deviation of the fitted Gaussian function (red solid line) is $0.46^\circ \pm 0.01^\circ$.}
    \label{fig:phasenoise}
\end{figure}
The measurement of the phase noise of Alice's homodyne detection during random switching between the $X$ and $P$ quadrature was performed by measuring the beat between the local oscillator and the bright control beam which was used to lock the squeezed-light sources.
Scanning the local oscillator's phase yielded a calibration between the measured output voltage of the homodyne detector's circuit and the phase angle between local oscillator and signal field.
Measurements were taken with an oscilloscope while randomly switching the quadrature. As for the quadrature measurements (see above) a segment of $1\,\mu$s was taken $3\,\mu$s after switching quadratures and the mean value was calculated.
Since the local oscillator was switched randomly between the $X$ and $P$ quadrature the phase noise is symmetric between the quadratures, hence $V_X = V_P$. Fig.~\ref{fig:phasenoise} shows a histogram of the phase noise measurement for $10^5$ samples. The red solid line shows a fit of a Gaussian distribution.
The standard deviation of the phase noise was determined to $(0.46 \pm 0.01)^\circ$ which is quite low despite the randomly switched quadrature angle~\cite{Mehmet2011}. 
Thereby the error was determined by bootstrapping $1000$ data points from a total of $10000$.

\subsection{Quantum Random Number Generator}

The security of the protocol relies on the use of true random numbers which are needed by Alice and Bob to choose between the $X$ and $P$ quadrature, and to determine a random hash function during privacy amplification. We implemented a quantum random number generator following a scheme of Ref.~\citenum{Gabriel2010} which is based on vacuum state measurements performed by a balanced homodyne detector. For this purpose we implemented another balanced homodyne detector with blocked signal port using an independent $6$\,mW $1550$\,nm beam from a fiber-laser as local oscillator. The output of the homodyne detector circuit was anti-alias filtered by a $50$\,MHz fourth-order Butterworth filter and sampled with a sampling frequency of $256$\,MHz by a data acquisition card. The data was subsequently mixed down digitally at $8$\,MHz, lowpass filtered with a $200$-tap finite-impulse-response filter with a cut-off frequency of $5$\,MHz and down-sampled to $2$\,MHz. The generation of the random numbers from the data stream followed the procedure in Ref.~\citenum{Gabriel2010}.

\subsection{Security Proof Considering Measurement Flaws}

We use the security proof from Ref.~\citenum{Furrer2012} and generalize it to phase errors in Alice's measurement of $X$ and $P$. It has been shown that if the protocol passes, a secure key of length~\cite{Furrer2012} 
\begin{equation}\label{keyrate0}
 \ell \le n ( \log \frac{1}{c(\delta)} - \log \gamma(d_\text{pe}^0) ) - \ell_{\text{LK} } - O( \log\frac{1}{\epsilon} )\ ,
\end{equation}
can be extracted. Here, $n=N-k$ is the number of samples used for the key generation, $\gamma$ is a bound on the correlation between Alice and Bob depending on the previously agreed average distance threshold $d_\text{pe}^0$ and $\ell_{\text{LK} }$ is the number of communicated bits in the error correction protocol. The only term depending on Alice's measurement device is $c(\delta)$, which refers to the overlap of the discretized $X$ and $P$ measurements performed by Alice. In the case of ideal $X$ and $P$ measurements satisfying the commutation relation $[X,P]=i\hbar$ one obtains an $c(\delta) \leq \delta^2/(2\pi\hbar)$, where equality holds approximately for relevant sizes of $\delta$. 

Let us now assume that due to experimental imperfections the actual measurements $X$ and $P$ deviate by a phase $\theta_X$ and $\theta_P$ from the ideal measurements, where $\theta_X$ and $\theta_P$ are distributed according to a Gaussian distribution with variance $V_X$ and $V_P$ centered at $0$. Then we find that $X$ and $P$ satisfy the canonical commutation relation $[X,P]=i\hbar'$ with $\hbar'=\hbar\cos\theta$, $\theta=\theta_X+\theta_P$. This then results in an overlap $c(\delta,\theta)= \delta^2/(2\pi\hbar') = c(\delta)/\cos\theta$. 

Considering now $n$ independent such measurements, we obtain
\begin{align} 
\log \prod_i \frac{\cos\theta_i }{c(\delta)}
 = n \log 1/c(\delta) + \sum_i \log \cos(\theta_i) \, .
\end{align}  
Using that $\log \cos(\theta) \geq  -\theta^2/(2\ln2)$, we can bound $ \sum_i \log \cos(\theta_i)\geq -  1/(2\ln2)\sum_i \theta_i^2$ and Hoeffding's inequality yields that $\sum_i \theta_i^2 \leq n (V_X+V_P + \varepsilon_P)$ with probability exponentially small in $\varepsilon_P^2 n$. Here we used that $\theta_X$ and $\theta_P$ are independent such that the expectation of $\theta^2$ is $V_X+V_P$.   
 Plugging this into~\eqref{keyrate0}, we find that for Gaussian phase noise with variance $V_X,V_P$ a secure key of length
\begin{equation}\label{keyrate}
\ell \le n ( \log \frac{1}{c(\delta)}-\frac{V_X+V_P}{2\ln2}  - \log \gamma(d_\text{pe}^0 ) ) - \ell_{\text{LK} } - O( \log\frac{1}{\epsilon} )\ 
\end{equation}
can be generated.

\subsection{Classical Post Processing}

The main post-processing is performed with the AIT QKD software. For the current protocol the following algorithms are combined: 
(i) the binning of the synchronized outcomes,
(ii) the estimation algorithm for CV QKD, 
(iii) the reconciliation algorithm for CV QKD, 
(iv) the confirmation algorithm, and
(v) the privacy amplification algorithm. 
All classical messages during the protocol are authenticated with a message authentication code using a pre shared secret key to select a random function from a set of (almost strongly two-universal) polynomial hash functions.

(i) First, Bob's samples in the $P$ quadrature are multiplied by $-1$ to account for the anti-correlation. Alice and Bob then discretize their sifted samples into $2^d$ bins of equal size $ \delta $ in the interval $[-\alpha, \alpha]$. The remaining outcomes associated to the intervals $(-\infty,-\alpha)$ and $(\alpha,\infty)$ are joined to $(-\alpha,-\alpha+\delta)$ and $(\alpha-\delta,\infty)$, respectively. The $2^d$ bins are identified with the key generation alphabet $ \alphakg = \{0,1\}^d $ and each bin (symbol) has a unique binary representation of $d$ bits. Alice and Bob obtain the binned sifted samples $X^{\text{sift}}_A \in \alphakg^{N}$ and $X^{\text{sift}}_B\in \alphakg^{N}$, respectively. Throughout the experiment we have used a key generation alphabet of size $\vert \chi_\text{kg} \vert =2^{12}$. 

(ii) In the estimation module for CV QKD the average distance between Alice's and Bob's binned symbols is estimated.  Alice chooses a random index set $\estiset \subset \{1,2,\dots,N\}$ of size $\vert \estiset \vert = k$ for estimation and communicates $\estiset$ together with the corresponding binned symbols $X_A^\text{pe}:=X^{\text{sift}}_A(\estiset)$ to Bob. Bob determines his corresponding binned raw key symbols $X_B^\text{pe}:=X^{\text{sift}}_B(\estiset)$, calculates the mean difference $d_\text{pe}$ between $X_A^\text{pe}$ and $X_B^\text{pe}$ (see~Eq.~(\ref{eqn:pe test})), and checks that $d_\text{pe} \leq d_\text{pe}^0$. Here, $d_\text{pe}^0$ has been determined before the run of the protocol by a theoretical estimation given the characterization of the source, the fiber loss and excess noise. If the test passes they continue with the protocol and both parties remove the $k$ estimation samples from their sifted samples to form their raw keys $X_A:=X^{\text{sift}}_A\setminus X_A^\text{pe} \in \alphakg^{N-k}$ and $X_B:=X^{\text{sift}}_B\setminus X_B^\text{pe} \in \alphakg^{N-k}$.

(iii) The reconciliation module for CV QKD implements the hybrid reconciliation protocol. As the security analysis uses direct reconciliation, Bob has to correct his raw key $X_B$ to match with Alice's $X_A$ to generate a common raw key $X$. The hybrid reconciliation used to correct Bob's noisy raw key operates directly on the key generation alphabet $ \alphakg $. In preparation for the hybrid reconciliation, two additional alphabets $\hi{\chi}$ and $\lo{\chi}$ are introduced such, that $ \chi_\text{kg} = \hi{\chi} \times \lo{\chi} $. Hence, each symbol $x \in \alphakg$ has a unique decomposition $x=(\hi{x},\lo{x})$ with $ \hi{x} \in \hi{\chi} $ and $\lo{x} \in \lo{\chi}$. We take for $\hi{x}$ the $d_2$ most significant bits of the binary representation of $x$, and for $\lo{x}$ the remaining $d_1=d-d_2$ least significant bits of the binary representation of $x$. We thus decompose the raw keys as $X=(\hi{X},\lo{X})$, where $\hi{X}$ and $\lo{X}$ denote the sequence of the $d_2$ most and the $d_1$ least significant bits of each key symbol, respectively.
The reconciliation module performs the following steps: 

(iii-a) Based on the variance of her binned raw key and the samples $X_A^\text{pe}$ and $X_B^\text{pe}$, Alice determines $d_1$, $d_2$, and the code rate $R$ such that the expected leakage is minimized w.r.t.~the entropy in Bob's symbols, and transmits these parameters to Bob. 

(iii-b) Then Alice communicates $\least_A$ to Bob who reconciles $\lo{X}_B$ simply by setting $ \least_B:=\least_A $. Hence, the errors which are left in Bob's key $X_B$ are reduced to the errors in $\most_B$. Non-binary LDPC reconciliation is used to correct $\hi{X}_B$ as described in the next step.

(iii-c) Both Alice and Bob split their $\hi{X}_A$ and $\hi{X}_B$ into blocks $\hi{X}_A^{(\ell)}$ and $\hi{X}_B^{(\ell)}$, $\ell=1,\dots, \frac{N-k}{n}$, each with $n=10^5$ elements of $\hi{\chi}$.
For this step we identify $ \hi{\chi}$ with GF($2^{d_2}$), the Galois field with $2^{d_2}$ elements. For each block $\most_A^{(\ell)}$, Alice uses the parity check matrix $H$ of an LDPC code over GF($2^{d_2}$) and rate $R$ to calculate the syndrome $s^{(\ell)}:=H\cdot \most_A^{(\ell)}$. Alice sends the syndrome $s^{(\ell)}$ to Bob. For all elements $j \in GF(2^{d_2})$ and for all indices $i\in\{1,\dots,n\}$ in the block Bob calculates the conditional probability that $(\most_A^{(\ell)})_i=j$, given that Bob has obtained $(\most_B^{(\ell)})_i$ and given Alice's value $(\least_A^{(\ell)})_i$. Bob uses these probabilities to initialize a non-binary belief propagation decoder.

The non-binary belief propagation decoder operates in the probability domain using the multi-dimensional Hadamard transform to speed up the check node operations~\cite{Barnault2003}. Using the syndrome $s^{(\ell)}$ and the conditional probabilities mentioned above, this decoder calculates Bob's estimate $\tilde{X}_A^{(\ell)}$ of Alice's block $X_A^{(\ell)}$.

We have constructed parity check matrices of non-binary LDPC codes over Galois fields of order 32, 64, 128, and 256 with code rates $R\in\{0.50, 0.51, \dots, 0.95\}$. Each LDPC code has a variable-node degree of two, is check-concentrated, and has a block length of $10^5$ symbols. We used the progressive edge-growth algorithm~\cite{Hu2005} to construct binary codes in a first step. Then each edge has been assigned a random non-zero element of the corresponding Galois field.\cite{Hu2005} Alice and Bob have access to all non-binary parity check matrices.

In our proof-of-principle experiment the error reconciliation step took about $2$\,h on a single CPU core for the largest data set of $2\times 10^8$ samples. Taking into account the about $30$\,min to measure the data, real-time error reconciliation could in principle be achieved by splitting the task to e.g.\ $5$ CPU cores. Alternatively, to speed up the computation LDPC decoder algorithms with reduced complexity could be employed~\cite{Voicila2010}.

(iv) After each block has been corrected, a confirmation step establishes the correctness of the protocol using a family $H$ of (almost) two-universal hash functions with $\text{Prob}_{h\in_R H}(h(x_1)=h(x_2))\le \epsilon_c$ for all $x_1\ne x_2$. For each block Alice chooses a hash function $h$ randomly from $H$ and communicates her choice to Bob. Alice and Bob apply this hash function to their blocks $X_A^{(\ell)}$ and $\tilde{X}_A^{(\ell)}$ and exchange the results. If their results agree the probability that Alice's and Bob's blocks are different is bounded from above by $\epsilon_c$. If their results disagree their blocks are definitely different, and they discard them. 

(v) Finally, Alice and Bob feed the sequence of all confirmed blocks into the privacy amplification module. Given the accumulated leakage $\ell_{\text{LK}}$ in bits from the previous protocol steps the secure key length $\ell$ is calculated according to equation~\eqref{keyrate}.
Alice chooses a hash function randomly from a two-universal hash family and communicates her choice to Bob. Then Alice and Bob both apply this hash function to the reconciled blocks and obtain the $\epsilon$-secure key $K_\text{sec}$.

\paragraph{\textbf{Acknowledgements}} \small This research was supported by the Deutsche Forschungsgemeinschaft (projects SCHN 757/5-1 and WE 1240/20-1), the Centre for Quantum Engineering and Space-Time Research (QUEST) and the Vienna Science and Technology Fund (WWTF) (project ICT10-067). TG and VH thank the IMPRS on Gravitational Wave Astronomy for support. TG also acknowledges support from the H.C.\ Ørsted postdoctoral programme. FF acknowledges support from Japan Society for the Promotion of Science (JSPS) by KAKENHI grant No. 24-02793. CP would like to thank Gottfried Lechner for very helpful conversations. RFW acknowledges support from the European network SIQS.

\paragraph{\textbf{Author Contributions}} \small TG and VH built the experimental setup with theory support from JD, FF and TF under supervision of RFW and RS. FF extended the security proof. JD, FF and CP developed the error reconciliation protocol and CP implemented and optimized it. TG and VH performed the experiment and TG analyzed the data with help from CP. TG, FF, CP and RS wrote the manuscript with contributions from all authors.

\paragraph{\textbf{Competing Financial Interests}} The authors declare no competing financial interests.

\end{document}